\documentclass[prl,aps,twocolumn,showpacs,floatfix]{revtex4}

\usepackage{epsfig}
\usepackage{amssymb}
\usepackage{amsmath}
\usepackage{amsfonts}
\usepackage{epstopdf}
\usepackage{verbatim}
\usepackage{color}
\usepackage[normalem]{ulem}
\begin{document}

\title {
 Low temperature Hall effect in bismuth chalcogenides thin films
}

\author{ A.\,Yu.~Kuntsevich$^{a,b}$\thanks {e-mail:
alexkun@lebedev.ru},A.A. Gabdullin$^c$, V.A. Prudkogliad$^a$, Yu.G. Selivanov$^a$, E.G. Chizhevskii$^a$, V.M. Pudalov$^{a, b}$}

\affiliation{$^a$P.\,N.~Lebedev Physics Institute, 119991 Moscow, Russia}

\affiliation{$^b$National Research University Higher School of Economics, Moscow 101000, Russia}
\affiliation{$^c$National Research Nuclear University MEPhI (Moscow Engineering Physics Institute), Kashirskoe sh. 31, 115409 Moscow, Russia}
\begin{abstract}

Bismuth chalcogenides are the most studied 3D topological insulators. As a rule, at low temperatures thin films of these materials demonstrate positive magnetoresistance due to weak antilocalization. Weak antilocalization should lead to resistivity decrease at low temperatures;
%which is in contrast  to the experimental observation.
in experiments, however, resistivity grows as temperature decreases.
 From transport measurements  for several thin films (with various carrier density, thickness, and carrier mobility), and by using purely phenomenological approach, with no microscopic theory, we show that the low temperature growth of the resistivity is accompanied by growth of the Hall coefficient, in agreement with diffusive electron-electron interaction correction mechanism. Our data reasonably explain the low-temperature resistivity upturn.

\end{abstract}
\maketitle
 Among numerous three-dimensional (3D) topological insulators, Bi$_2$Se$_3$ and Bi$_2$Te$_3$ are the most studied materials.
They have a large
band gap in the bulk, ~300\,meV and ~150\,meV respectively, and can be rather easily synthesized.

Revealing  topological and spin related properties in these systems
as a rule requires reduced temperatures. Lowering the temperature, in its turn, not merely simplifies the system, but leads to new questions. One of them is the so-called low temperature transport paradox \cite{LuShen}: almost all films in weak fields demonstrate positive magnetoresistance due to suppression of the weak antilocalization (WAL) correction. The WAL correction should lead to decrease of resistivity as temperature decreases  ($d\rho/dT  >0$, i.e. ``metallic'' temperature dependence). Contrary to this expectation, almost all films, grown by various methods\cite{example1,example2,magnetron} and crystalline flakes split off the bulk materials \cite{flakesexample1} at low temperatures demonstrate ``insulating''-type dependence  ($d\rho/dT<0$) \cite{Rev1}. This observations mean that some other mechanism exists on top of the weak antilocalization and drives the system towards insulator as temperature decreases.

In Refs.\cite{magnetron, flakesexample1,LiuEEEvidence, WangEEEvidence, TakagakiEEEvidence,DeiEEEvidence, LuShen,LiuACSNano} it  was  conjectured that electron-electron interaction correction to conductivity $\Delta\sigma_{ee}$ could be a reason for such insulating behavior. In order to substantiate their point the authors in Refs.~\cite{magnetron, flakesexample1,LiuEEEvidence, WangEEEvidence, TakagakiEEEvidence,DeiEEEvidence, LuShen,LiuACSNano} refer to the microscopic electron-electron interaction(EEI) theory\cite{AltArLee,LeeRamakrishnan}.
Several groups attempted to fit their transport data with EEI theory, using the Fermi-liquid coupling constant $K_{\rm ee}\equiv 2\pi^2\hbar/e^2\times d(\Delta\sigma_{ee})/d\ln (T)$  as a fitting parameter; however, the latter values extracted  from fittings appeared to be very much scattered even for the samples with nominally similar carrier density.

Experimentally, the conjecture about quantitative  contribution of EEI to the transport does not have a  solid ground. Indeed, the EEI correction so far was extracted only  in one manner: by assuming that all low-$T$ transport properties are determined by only two effects, i.e. WAL and EEI. It was assumed that subtraction of the  theoretically calculated  WAL contribution leaves temperature dependence of conductivity only related to the EEI correction. This assumption, however, has not been checked independently. From the theoretical point of view, all TI thin films are multi-component systems, and as it was shown e.g. in Refs. \cite{LuShen,konig}, generalization of the electron-electron interaction correction on multi-component system is not reduced to simple results of Ref.~\cite{AltArLee}. Moreover, taking into account scattering between the subsystems further complicates the problem\cite{punnoose, kuntsevich_PRB_2007}. Therefore, from fitting of the temperature dependence of the resistivity solely, one can not rule out other mechanisms on top of the  WAL and EEI correction (such as, e.g., density-of-states correction \cite{doscorrection},  Maki-Thompson correction \cite{mtcorrection}, etc). In order to overcome this problem we use a theoretically substantiated fully phenomenological approach, that has already been experimentally tested for disordered two dimensional systems with a simple spectrum \cite{minkoveefirst}.

In this paper we measure not only the temperature dependence of resistivity, but also of the Hall resistivity. We found that the low temperature growth of the resistivity is accompanied by growth of the Hall coefficient, in agreement with diffusive electron-electron interaction correction mechanism \cite{altshulerreview, altshulerhall}.  We discuss only phenomenology of the EEI correction and do not address its microscopic structure. We compare two methods to determine $K_{\rm ee}$: (i) from temperature dependence of the resistivity $K_{\rm ee}^{xx}$, similarly to the previous investigators\cite{magnetron,flakesexample1,LiuEEEvidence, WangEEEvidence, TakagakiEEEvidence,DeiEEEvidence, LiuACSNano} and (ii) from temperature dependence of the Hall resistivity $K_{\rm ee}^{xy}$. We show that in some samples $K_{\rm ee}^{xx}$ and $K_{\rm ee}^{xy}$ coincide, whereas in some other samples these two values deviate from each other.

\section{Theoretical background}

The idea to determine $K_{\rm ee}$ from $T$-dependence of the Hall resistance
 goes back to 1980-s, and originates from the unique property (that is independent of dimensionality) of the diffusive electron-electron interaction correction  not to affect  the Hall conductivity\cite {altshulerhall, altshulerreview}. In other words, for the magnetoconductivity tensor in perpendicular magnetic field one has:
\begin{equation}
\sigma=\sigma_D+\left( \begin{array}{cc}
\Delta\sigma_{\rm ee}  & 0 \\
0  & \Delta\sigma_{\rm ee} \\
\end{array} \right)
\label{property}
\end{equation}

The first term, Drude conductivity for single-component two dimensional system can be written as:
\begin{equation}
\sigma_D=\frac{ne\mu}{1+\mu^2 B^2} \left( \begin{array}{cc}
 1  & \mu B \\
-\mu B  & 1 \\
\end{array} \right)
\end{equation}

Inverting the conductivity tensor, and assuming that $\Delta\sigma_{\rm ee} \ll ne\mu$, we obtain for the resistivity tensor:
\begin{equation}
 \rho \approx \frac{1}{ne\mu} \left( \begin{array}{cc}
 1  & -\mu B \\
\mu B  & 1 \\
\end{array} \right)
-\frac{\Delta\sigma_{\rm ee}}{(ne\mu)^2}\left( \begin{array}{cc}
 1-\mu^2 B^2    & -2\mu B \\
2\mu B   & 1-\mu^2B^2 \\
\end{array} \right)
\label{aamrtensor}
\end{equation}

In the limit of low magnetic fields $\mu B\ll 1$ we come to the famous relation \cite{altshulerreview}:
\begin{equation}
\frac{\Delta\rho_{xy}}{\rho_{xy}}=2 \frac{\Delta\rho_{xx}}{\rho_{xx}}=-2 \frac{\Delta\sigma_{\rm ee}}{\sigma_D}
\label{maineq}
\end{equation}

\begin{figure}[t]
\begin{center}
\centerline{\psfig{figure=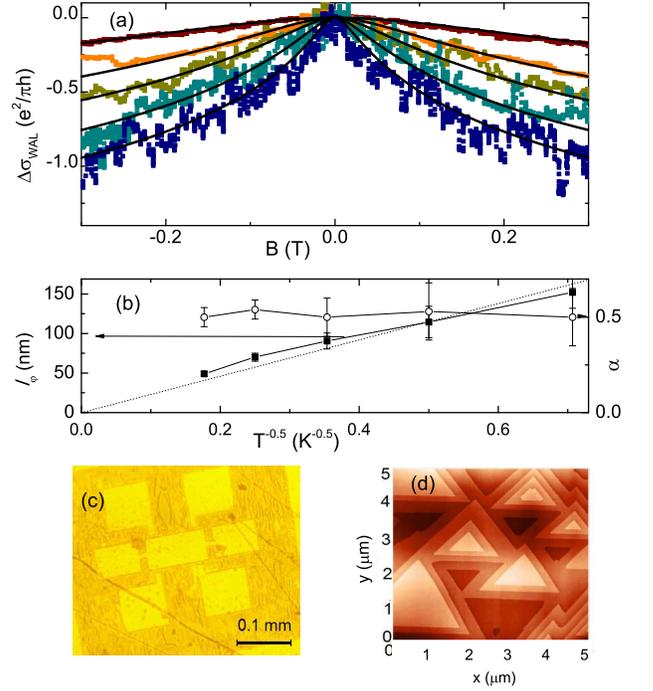, width=240pt}}
 \caption{(Color online) (a) Magnetoconductance  in $B_\perp$ field, due to weak antilocalization, for sample Bi$_2$Se$_3$-685  (dots). Temperatures from top to bottom are 32, 16, 8, 4 and 2\,K, respectively. Black lines show fit with Eq.~(\ref{HLN}). (b) The corresponding temperature dependencies of the fitting parameters $l_\phi$ (left axis, filled boxes) and $\alpha$ (right axis, open circles) in Eq.~(\ref{HLN}). (c) Optical image of the mesa for sample Bi$_2$Se$_3$-685. (d) Typical AFM image of the epitaxial Bi$_2$Se$_3$ film before lithography. Step heights correspond to one quint-layer (about 1 nm).
 }
\label{fig3}
\end{center}
\end{figure}

In practice, for topological insulator thin films, the top and bottom conductive channels, and the bulk are connected in parallel and the Drude conductivity tensor is multi-component; however, this fact does not affect the low-field limit (see Appendix I). Since it is very hard to achieve  mobilities exceeding 0.5 m$^2$V$^{-1}$s$^{-1}$ in epitaxial films of bismuth chalcogenides \cite{QHEBiSe1}, for most of the films the low-field linear Hall effect regime ($B<1/\mu$) extends to several Tesla. Another important fact is that WAL does not affect the Hall resistivity \cite{altshulerreview,fukuyama} , i.e. one can determine EEI contribution straightforwardly by using Eq.~(\ref{maineq}), provided the temperature dependence of the Hall effect originates from EEI-correction solely.

 Correspondingly, we exploit two ways to determine $K_{ee}$ experimentally:
(i) From the low field magnetoresistance we determine the amplitude of the WAL correction  and subtract it from the low-temperature $\sigma(T)$ dependence. The resultant temperature dependence of the resistivity is believed to be $\Delta\sigma_{ee}= K_{\rm ee}^{xx} e^2/(2\pi^2\hbar) \ln (T)$.
(ii) We analyze temperature dependence of the Hall resistance, using  Eq.~(\ref{maineq}). Assuming that correction is much smaller than the Drude value of the resistivity, we get $\Delta\sigma_{ee}(T)=\sigma_{D}\times [\rho_{xy}(B,T)-\rho_{xy}^{D}(B)]/\rho_{xy}^{D}(B)$. Here the Drude value of the Hall resistance $\rho_{xy}^D(B)$ is taken at
%the
elevated temperatures, $T\tau>1$ beyond the diffusive interaction regime. Since variation of the Hall resistivity with temperature is small, the particular choice of $\rho_{xy}^D$ is not important. Moreover, as we can see from our data at low enough fields, thus calculated $\Delta\sigma_{ee}(T)$ is $B-$independent.

\begin{figure}[t]
\begin{center}
\centerline{\psfig{figure=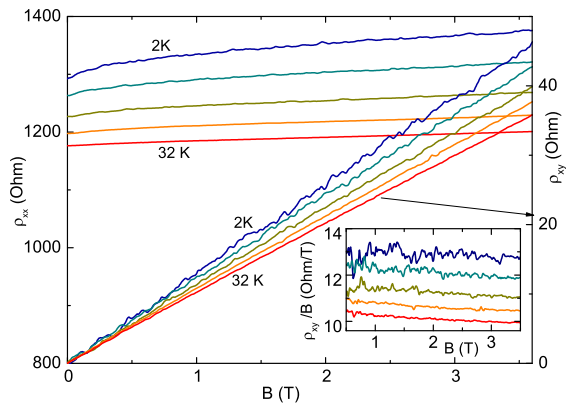, width=220pt}}
 \caption{(Color online) $B_\perp$ field magnetoresistivity (left axis) and Hall resistivity (right axis) for sample Bi$_2$Se$_3$-685 at temperatures 2, 4, 8, 16, and 32\,K (indicated on the panel).  The inset  shows Hall coefficient $\rho_{xy}/B$ versus magnetic field for the same data set.}
\label{fig1}
\end{center}
\end{figure}

\section{Experimental}

We studied several samples with various mobilities, and carrier densities, their main parameters are summarized in Table \ref{summarytable}.
Epitaxial Bi$_2S$e$_3$ and Bi$_2$Te$_3$ layers were grown on (111)-oriented BaF$_2$  substrates similarly to Refs.\cite{caha,hoefer}
(Bi$_2$Se$_3$-707 sample was grown on an Al$_2$O$_3$ substrate) by molecular beam epitaxy using a compound bismuth selenide/telluride effusion cell and an additional selenium/tellurium cell, respectively.  Corresponding fluxes were calibrated before and after the growth using an ion gauge beam flux monitor. The beam equivalent pressure flux  ratio of  Se:Bi$_2$Se$_3$ (Te:Bi$_2$Te$_3$) was maintained to be 2:1 for the stoichiometry control. The film  growth was performed at about 320\,C at the 0.3\,nm/min growth rate. Streaky  RHEED patterns during the growth evidenced of single crystalline films with a smooth surface.  After growth, the films on the substrate were cooled down to room temperature and 30\,nm Se or 100\,nm BaF$_2$  protective cap layers were deposited on the surface of the film or the films remained  unprotected (see Table \ref{summarytable}). X-ray diffraction scans show only the $3l$ allowed peaks indicating clear c-axis orientation in all the grown films. For film thicknesses ranging from 9 to 50\,nm, the full width at half maximum of the (0006) rocking curve does not exceed 0.1$^\circ$ confirming high crystalline quality of the samples. Thickness of the films was determined from (0006) and (00015) Bragg peaks thickness fringes as well as from atomic force microscope measurements (see example in Fig.\ref{fig3}d).

The studied samples were either defined lithographically as Hall bars (using Laser beam lithography system Heidelberg $\mu$Pg101) and mesa-etched  in oxygen plasma (see optical image in Fig.\ref{fig3}c) or manually cut with a razor blade. Manual scratching makes geometry of the sample (see error bars in Table \ref{summarytable}) not so well-defined, however it helps to preserve the active layer intact. The  leads to the contact pads were attached with either graphite or silver paint.  The results on both scratched and etched samples are similar. Charge transport in the most disordered Bi$_2$Se$_3$ samples showed up noise, increasing with lowering temperature (see data for sample Bi$_2$Se$_3$-685 below). Resistivity of all samples did not exceed 1.5 kOhm per square, so they all can be treated as ``good'' metals  where charge transport is slightly affected by quantum corrections.

>From the electronic transport point of view our thin films are quasi-two dimensional systems, i.e. they contain at least three interacting subsystems (two surfaces and the bulk), connected in parallel. We consider the system as two-dimensional, i.e.  throughout the paper, from the measured resistance $R$ [in Ohms], we calculate resistance per square $\rho= R(w/l)$ [in Ohms per square] and call it, for simplicity,   ``resistivity''; we further discuss corrections to its inverse value, the 2D conductivity. We believe that we can do so, because (i) in the low temperature limit the  out-of-plane motion of carriers is coherent (the phase breaking length presented below much exceeds the film thickness), (ii) resistivity of our thin films very weakly responds to the parallel magnetic field, and (iii) logarithmic corrections to conductivity, observed by us and by other investigators in similar films \cite{magnetron,TakagakiEEEvidence,WangEEEvidence,LiuEEEvidence} are intrinsic to two dimensional systems solely.

\begin{figure}[t]
\begin{center}
\centerline{\psfig{figure=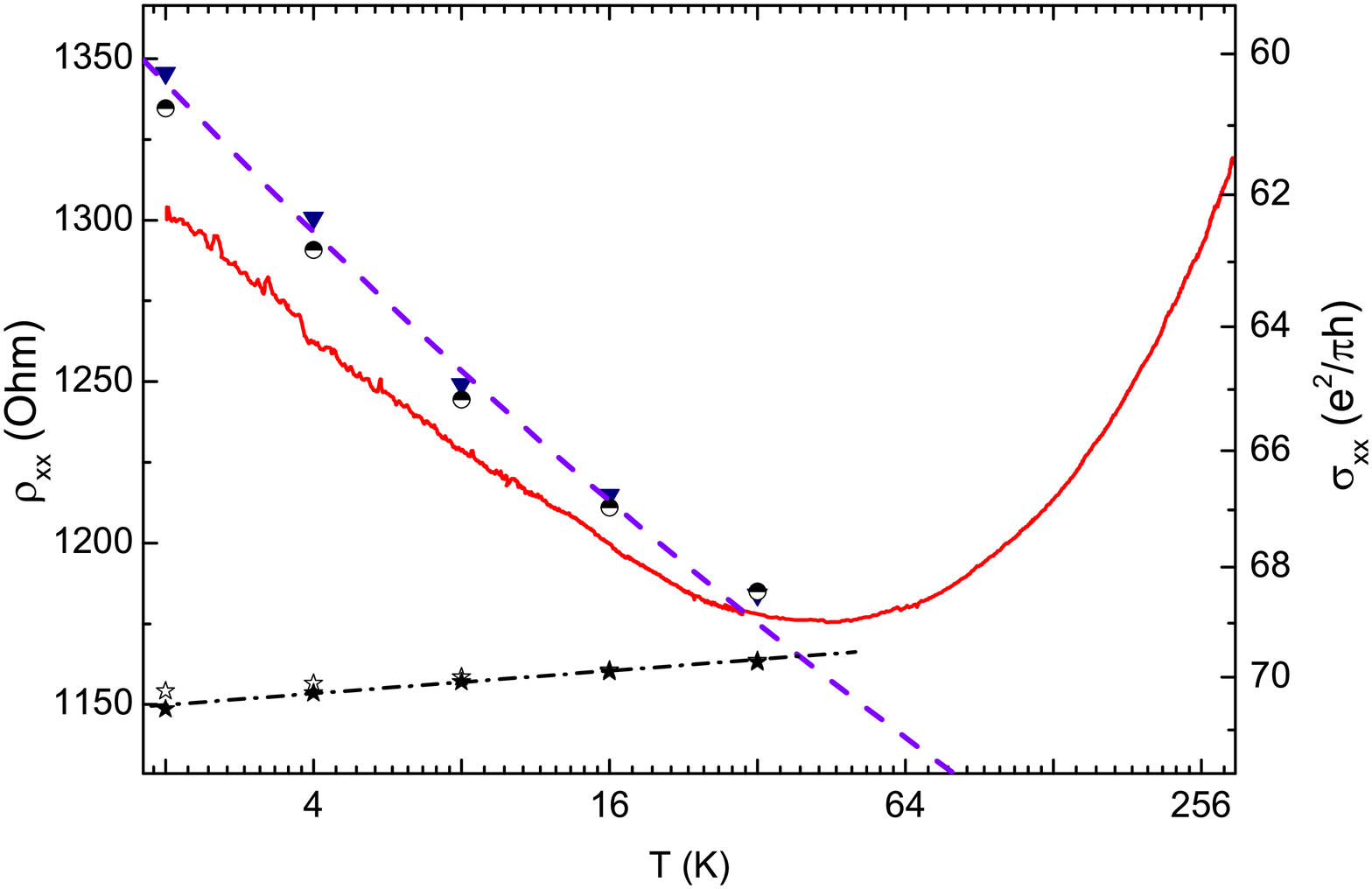, width=220pt}}
 \caption{(Color online) Temperature dependence of the resistivity/conductivity(left/right axis) for sample Bi$_2$Se$_3$-685 -- red curve. Symbols denote quantum corrections: triangles are from EEI, determined from the Hall slope according to Eq.~(\protect\ref{maineq}), circles -- temperature dependence of the resistivity minus WAL contribution; black stars -- WAL correction, determined from magnetoresistance  using Eqs.~(\ref{HLN}) and (\ref{RvsTWL}),  empty stars -- $T$-dependence of the WAL correction determined as $\Delta\sigma_{WAL}=\rho(B=0)^{-1}-\rho(B=1{\rm T})^{-1}$. Blue dashed line  shows theoretical fit to determine K$_{ee}$.}
\label{fig2}
\end{center}
\end{figure}

Resistivity and Hall resistance were measured using standard lock-in amplifier ($f=13-33$Hz) at temperatures 0.3-300\,K and in magnetic fields up to 7 Tesla. We used PPMS-9\,Tesla and  Cryogenic-21\,Tesla systems. Measurement current was chosen in the range 0.1-2 $\mu$A to ensure the absence of overheating at the lowest temperature. In order to compensate contact asymmetry, magnetic field was swept from  positive to negative values and the  $\rho_{xx}$($\rho_{xy}$)  data were (anti)symmetrized. Weak antilocalizaton was studied in separate low-field sweeps.

\section{Results}

Weak field magnetoresistance due to the weak antilocalization is typically processed using the simplified Hikami-Larkin-Nagaoka(HLN) formula:
\begin{equation}
\frac{\Delta\rho_{\rm WAL}(B)}{\rho_D^2}=\alpha\frac{e^2}{2\pi^2\hbar}\left [ \Psi\left(\frac{\hbar}{4el^2_\phi B} +\frac{1}{2}\right )-\ln \left( \frac{\hbar}{4el^2_\phi B}\right )\right  ]
\label{HLN}
\end{equation}

\begin{figure}[t]
\begin{center}
\centerline{\psfig{figure=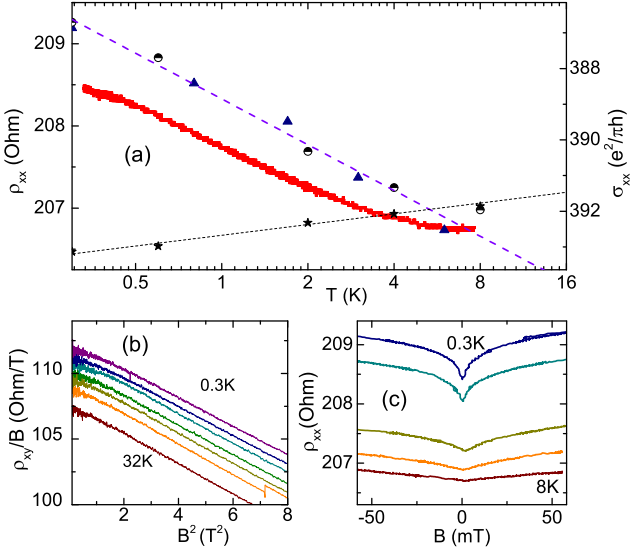, width=240pt}}
 \caption{(Color online) (a) Temperature dependence of the resistivity/conductivity (left/right axis) for sample Bi$_2$Se$_3$-724 at low temperatures -- red curve. Symbols show quantum corrections: triangles --  EEI, determined from the Hall slope according to Eq.~(\protect\ref{maineq}), circles -- temperature dependence of the resistivity minus WAL contribution, black stars --  WAL determined from magnetoresistance using Eqs.~(\protect\ref{HLN}) and (\ref{RvsTWL}),  empty stars -- $T$-dependence of the WAL correction determined as $R(B=0)-R(B=50{\rm mT})$. Dashed line -- fit to determine $K_{\rm ee}$. (b) Hall slope  versus magnetic field  for the same sample at various temperatures, (c) low-field magnetoresistance due to the weak antilocalization. }
\label{fig724}
\end{center}
\end{figure}

\begin{figure}[t]
\begin{center}
\centerline{\psfig{figure=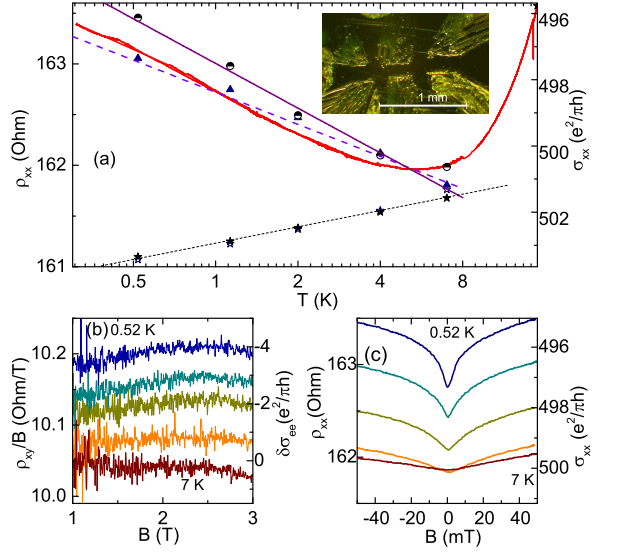, width=220pt}}
 \caption{(Color online) (a) Temperature dependence of the resistivity/conductivity (left/right axis) for sample Bi$_2$Te$_3$-677 at low temperatures -- red curve. Symbols are the quantum corrections: triangles -- from EEI, determined from the Hall slope according to Eq.~(\ref{maineq}), circles -- temperature dependence of the resistivity minus WAL contribution;  black stars -- temperature dependence of the WAL contribution determined from magnetoresistance  using Eqs.~(\ref{HLN}) and (\ref{RvsTWL}). Empty stars -- $T$-dependence of WAL determined as $R(B=0)-R(B=50{\rm mT})$.  Dashed blue line $\propto \ln (T)$ is a fit to determine $K_{\rm ee}^{xy}$ , solid violet line -- $\propto \ln (T)$ is a fit to determine $K_{\rm ee}^{xx}$.  Optical image of the sample is also indicated in the main panel. (b) Hall slope  versus magnetic field for the same sample at various temperatures, (c) Low-field magnetoresistance due to the weak antilocalization. }
\label{fig4}
\end{center}
\end{figure}

Here $l_\phi$ is the phase coherence length and $\alpha$ is the prefactor, i.e. adjustable parameter that denotes a number of WAL channels. This formula is applicable in the so-called diffusive transport regime, i.e. when both $l_\phi$ and magnetic length $l_B\equiv \sqrt{\hbar/eB}$ much exceed the mean free path $l$.
Figure \ref{fig3} shows examples of magnetoconductance due to WAL  for sample Bi$_2$Se$_3-$685 and their fits with HLN formula. Similarly to Refs. \cite{magnetron, DeiEEEvidence,Chenwalprefactor, Taskin}, we find that the prefactor $\alpha$ is close to 0.5 and does not demonstrate temperature dependence (average prefactors are summarized in Table \ref{summarytable}), indicating strong spin-orbit coupling. The phase coherence length is roughly proportional to $T^{-0.5}$, as it should be for dephasing caused by  electron-electron scattering \cite{altshulerreview}, similarly to observations of Refs.~\cite{magnetron, Chenwalprefactor, Taskin, flakesexample1},
The zero-field temperature dependence of the resistivity due to WAL is expressed as:
\begin{equation}
\frac{\Delta\rho_{\rm WAL}(T)}{\rho_D^2}=-\alpha\frac{e^2}{\pi^2\hbar}\ln(l_\phi/l)
\label{RvsTWL}
\end{equation}
 The amplitude of thus determined WAL correction is shown by black stars in Figs.~\ref{fig2}, \ref{fig724}a, \ref{fig4}a. Alternatively, WAL contribution might be estimated similarly to Refs.\cite{LiuACSNano,TakagakiEEEvidence} as a difference of conductivities at $B=0$ and at some elevated magnetic field $B_0$, where $T-$dependence of the WAL is already suppressed: $\Delta\sigma_{WAL}\approx\sigma(B=0)-\sigma(B_0)$. Interestingly, the results obtained by this method (empty stars in Figs.\ref{fig1}, \ref{fig724}a, \ref{fig4}a) with reasonable precision coincide with those obtained from data fitting with Eqs.~(\ref{HLN}) and (\ref{RvsTWL}).

\begin{table*}
\caption{Summary of sample parameters.\label{summarytable}}
\begin{tabular}{|c|c|c|c|c|c|c|c|c|c|c|}
  \hline
    Sample,& n,$10^{12}$ & $\mu$,cm$^2$&$\rho_D$,& $L$ &$w$& $d$ &$\alpha$&$K_{\rm ee}^{xy}$&$K_{\rm ee}^{xx}$& Comment\\
   & cm$^{-2}$ & V$^{-1}$s$^{-1}$&Ohm/$\Box$&$\mu$m&$\mu$m&nm&&&&\\  \hline
  Bi$_2$Se$_3$-724 & 5.9& 5000$\pm 500$ &205$\pm 20$&560$\pm$50&165$\pm$5&10&0.33$\pm 0.03$ &1.6$\pm0.15$&1.6$\pm0.15$& Se-covered, scratched\\\hline
  Bi$_2$Se$_3$-685 & 61 &  85 &1170&125&80&15& 0.5$\pm0.15$&3.1$\pm$0.3&3.1$\pm$0.3& mesa-etched\\\hline
  Bi$_2$Se$_3$-691 &140  & 37  &1216&120&75&18&0.5$\pm 0.1$ &2.5$\pm 0.3$&2.8$\pm0.3$& unprotected, mesa-etched\\\hline
   Bi$_2$Te$_3$-677 & 62 &  630 & 160&460$\pm 50$&125$\pm5$&23&0.7$\pm0.1$ &1.4$\pm 0.15$&1.95$\pm 0.2$& 12 nm BaF$_2$-covered, scratched\\\hline
      Bi$_2$Se$_3$-707 & 61 &  410 & 245&770$\pm 80$&195$\pm10$&17&0.65$\pm0.1$ &1.4$\pm0.2$&3.25$\pm0.4$& Al$_2$O$_3$ substrate, scratched\\\hline
%   Bi$_2$Se$_3$-711$^*$ & x &  x$\times 4.9$ &x$\times 0.204$x&1030$\pm100$&210$\pm 10$&20&-0.5 && $Se$-covered, scratched\\\hline
%   Bi$_2$Se$_3$-722$^*$ & x &  x$\times l/w$ &x$\times w/l$x&&37(42)&&-0.5 && $Se$-covered, scratched\\\hline
\end{tabular}
\end{table*}

The effect of EEI correction for the low  mobility mesa-etched sample Bi$_2$Se$_3$-685 is the most pronounced: Figure~\ref{fig1} shows magnetoresistance and Hall effect versus magnetic field in a wide range of temperatures. As magnetic field increases, there is a weak positive temperature-independent MR on top of WAL. This MR is a consequence of several parallel conducting channels with various mobilities (bulk and surfaces,  see Appendix I). Hall resistance  is linear-in-$B$ and has an interesting feature:  the Hall slope grows with  decrease of temperature, as it is clearly seen from Fig.~\ref{fig1}. Such huge variation  of the Hall slope with temperature ($\sim 20\%$ - see insert to Fig.~\ref{fig1}) would be prohibited by overall electro-neutrality of the system  were the Hall resistance  related to the inverse carrier density; however it is possible that the EEI correction comes into play and should be added to the Drude conductivity, as  Eq.~(\ref{aamrtensor}) suggests.

If we interpret the temperature-dependent part of the Hall slope within Eq.~(\ref{maineq}) [for more details, see section ``Theoretical background'', method (ii)], we directly get temperature dependence of the EEI correction
and may use it for comparison with the corresponding variation of the resistivity
( triangles in Fig. \ref{fig2}). When  we subtract WAL (stars in Fig.~\ref{fig1}) from  the $\rho_{xx}(T)$ dependence, we get almost indistinguishable temperature dependence of the resistivity (see circles in Fig.~\ref{fig1}).  The observed agreement is the one of the central results of our paper. Moreover, this temperature dependence is visibly logarithmic, $\Delta\sigma_{\rm ee}=K_{\rm ee}\times e^2/(2\pi^2\hbar)\ln (T)$, and our observation means that $K_{\rm ee}^{xx}=K_{\rm ee}^{xy}$.

The value of $K_{\rm ee}=3.1$ is enormously large and exceeds those reported in Refs.~\cite{LiuEEEvidence, WangEEEvidence, TakagakiEEEvidence,DeiEEEvidence,LiuACSNano}.  Moreover, even within the theoretical predictions \cite{LuShen} there is no way to obtain $K_{\rm ee}$ more than 2.  A possible physical origin of the enhanced $K_{ee}$ could be effective decoupling of conductive channels. This mechanism was suggested in Ref.~\cite{magnetron}, where $K_{ee}>2$ was also observed.  Indeed, for such high electron densities, many quantization subbands are filled. If all these subbands generated independent conductive channels, each of them should have its own EEI correction to conductivity $\Delta\sigma_{ee}^i$, with $K_{ee}^i\sim 1$. In this case the total EEI correction should be a sum of all contributions and  may be large. In practice, however,  electrons from different subbands interact with each other and the amount of  singlet terms may be reduced to one. The question why in this particular sample Bi$_2$Se$_3$-685 (and probably Bi$_2$Se$_3$-691, see Table \ref{summarytable}) such interaction is suppressed remains opened.

Importantly, the coincidence of $K_{\rm ee}^{xx}$ and $K_{\rm ee}^{xy}$ is not occasional, rather, it is observed in several samples (see Table \ref{summarytable}). For example, sample Bi$_2$Se$_3$-724 (see Fig. \ref{fig724}) contains high mobility (presumably surface) carriers atop of low-mobility (presumably bulk) carriers. As a result, its Hall coefficient visibly  drops with field (see Appendix I) starting from already $\sim 1$\,Tesla (see Fig. \ref{fig724}b). Nevertheless, temperature dependence of the Hall effect is easily seen,  its relative variation $\Delta\rho_{xy}/\rho_{xy}\equiv [\rho_{xy}(B_0,T)-\rho_{xy}(B_0,T_0)]/\rho_{xy}$ is almost $B_0$-independent (here $B_0$ is the magnetic field at which Hall effect is analyzed), and we again get $K_{\rm ee}^{xx}\approx K_{\rm ee}^{xy}$.

The prefactor $\alpha$ has a reasonable value of about 0.33. Similar values of $\alpha$ were also reported in Refs. \cite{magnetron,TakagakiEEEvidence}. The value of $K_{\rm ee}=1.6$ is also comparable with those observed in Ref. \cite{TakagakiEEEvidence}. Both non-ideal $K_{\rm ee}$ and $\alpha$ values were explained theoretically in Ref.~\cite{LuShen}. Prefactor $\alpha$ might also differ from 0.5 because the effective $w/l$ ratio for sample made by scratching, might be determined only roughly; however within our phenomenological approach this uncertainty does not modify the overall result (see discussion below and Appendix II).

We note that the coincidence of $K_{\rm ee}^{xx}$ and $K_{\rm ee}^{xy}$ is not observed universally in all samples. For example,
Fig.~\ref{fig4} shows temperature dependence of the resistivity, WAL, and Hall effect in bismuth telluride sample  Bi$_2$Te$_3$-677 with the same carrier density as Bi$_2$Se$_3$-685 and an order of magnitude higher mobility. In bismuth telluride, the phonon-scattering-limited metallic temperature dependence of the resistivity is much stronger and extends to much lower temperatures.

Correction to the Hall coefficient is logarithmic in temperature (see Fig. \ref{fig4}b recalculated using Eq.~(\ref{maineq}) to blue triangles in Fig.\ref{fig4}a), but this temperature dependence is even weaker than the temperature dependence of resistivity. If we subtract WAL from the resistivity, we get circles in Fig. \ref{fig4}a with stronger temperature dependence. In other words $K_{\rm ee}^{xx}$ is about factor of two larger than $K_{\rm ee}^{xy}$. Possible reasons for  the difference are discussed below. At the same time, the WAL prefactor in this sample has a reasonable and $T$-independent value of 0.7 (see Fig. \ref{appendix}c). Bi$_2$Se$_3$-707 sample is pretty much similar to Bi$_2$Te$_3$-677.

\section{Discussion}
Importantly, inexact definition of the sample geometry by scratching (see example in the inset to Fig.~\ref{fig4}a) does not affect the results obtained within our phenomenological approach. Indeed, according to Eq.~\ref{maineq}, correction to the Hall coefficient is recalculated directly to the correction to diagonal resistance. When evaluating WAL contribution we measure prefactor $\alpha$ using Eq.~\ref{HLN} and then substitute the result with the same prefactor $\alpha$ into Eq.\ref{RvsTWL}. If the geometrical factor ($w/L$ in case of the Hall bar geometry) is calculated incorrectly, we automatically get the modified value of $\alpha$, but the total contribution of WAL in temperature dependence of the resistivity, calculated using Eq. \ref{RvsTWL} will remain correct. Observation of $\alpha\approx 0.5$ in almost all cases means that we are not mistaken with $w/L$ ratio.

Remarkable property of both WAL and EEI corrections is their logarithmic temperature dependence.  This logarithm has a high temperature cut off when the corresponding length (interaction or phase breaking) becomes comparable with the mean free path (in our samples this equality corresponds to 10-300\,K, i.e. above the resistivity minimum temperature). In this low temperature limit we neglect temperature dependence of the mean free time.
 The functional dependence of the result for the amplitude of both WAL corrections to conductivity $\Delta\sigma_{WAL}$ and EEI correction $\Delta\sigma_{\rm ee}$ is similar:  the amplitude is roughly proportional to (i)the corresponding amplitude factor of the order of unity ($\alpha$ in case of WAL, and Fermi-liquid coupling dependent term $K_{\rm ee}$ in case of EEI), (ii) $\ln (T\tau)$ \cite{altshulerreview, minkoveefirst} in case of EEI (where $\tau$ is the mean free time), and $\ln (\tau_\phi/\tau)$ in case of WAL.  $\tau_\phi$ is usually inversely proportional to temperature, therefore we get the same dependence. Since $\Delta\rho=-\rho_D^2\Delta\sigma\propto\rho_D^2\ln (\rho_D)$, the main parameter that governs the amplitude of the correction to resistivity is the sheet resistance value. Indeed, in Fig.~\ref{fig1} for Bi$_2$Se$_3$-685 with resistivity about 1 kOhm per square we see huge variation of the resistivity with temperature (about 10\% i.e. 100 Ohm), whereas for Bi$_2$Se$_3$-724 with resistivity about 200 Ohm per square (see Fig. \ref{fig724}) the overall variation of  resistivity is about 0.5\%, i.e. 1 Ohm.

It should be discussed why the data for samples Bi$_2$Te$_3$-677 and Bi$_2$Se$_3$-707 deviate from those for the other bismuth selenides.
Naively, the difference between Bi$_2$Te$_3$ and Bi$_2$Se$_3$ thin films is as follows:
(i) selenides are n-doped, whereas sample Bi$_2$Te$_3$-677 is p-doped;
(ii) spectra of carriers in Bi$_2$Te$_3$ are hexagonally wrapped. This wrapping was shown to modify the ordinary low-temperature resistivity behavior \cite{maslov}. It is not clear whether the wrapping somehow affects EEI correction to conductivity.
Close similarity between Bi$_2$Te$_3$-677 and Bi$_2$Se$_3$-707 samples does not support the latter possibility.

The agreement between resistivity and Hall resistivity in these samples would become much better if instead of Eq.~(\ref{maineq}) one would use $\Delta\rho_{\rm xx}/\rho_{\rm xx}=\Delta\rho_{\rm xy}/\rho_{\rm xy}$. Such functional temperature dependence of the Hall effect might originate from variable number of carriers; this mechanism was suggested, e.g., in Refs. \cite{HeJAP, kuntsevich_JETPL_2005}.  Indeed, if only one group of high mobility carriers determines transport, then $\rho_{xx}=(ne\mu)^{-1}$, and $\rho_{xy}=B\cdot(ne)^{-1}$. Variation of $n$ in this case leads to $\Delta\rho_{xx}/\rho_{xx}=\Delta\rho_{xy}/\rho_{xy}$, contrary to Eq.~(\ref{maineq}).
Nevertheless, we think
that this mechanism is irrelevant here, because we clearly observe logarithmic temperature dependence of the Hall effect, instead of the expected exponential activated dependence \cite{HeJAP, kuntsevich_JETPL_2005}.
%Moreover,
%it would be unphysical
%there is no reason to believe that in samples 685, 724 and 691 the number of carriers stays fixed independently of temperature, whereas for samples 677 and 707 it changes. As it is seen from the insert to Fig. \ref{fig1}, the Hall coefficient for sample Bi$_2$Se$_3$-685 grows by 30\% as temperature decreases from 32 to 2 K. It is very hard to imagine a mechanism that can store such a huge amount of carriers ($\sim2\cdot 10^{13}$ cm$^{-2}$) at temperatures well below all relevant energy scales (band gap, Fermi energy,  etc).

In fact, the lack of quantitative agreement with quantum correction theory is not surprising  because of the complexity of the studied films, and was observed several times even for systems with much simpler spectrum \cite{minkoveefirst,kuntsevichee}.
 As a possible clue to the explanation of this discrepancy, the prefactor $\alpha >0.5$ was found to be unexpectedly high in samples Bi$_2$Te$_3$-677 and Bi$_2$Se$_3$-70. Theoretically (see \cite{LuShen} and references therein) one expects $-0.5<\alpha <0.5$. Observation of the anomalous prefactor value out of this interval means that the sample is more complicated than the naive model presumes.
There are many unexplored factors, whose effect on transport properties of bismuth chalcogenides is not yet explored, e.g. formation of Bi- bilayers \cite{bibilayers}, spatial inhomogeneity, surface oxidation and irreproducible band-bending \cite{shklov}, intersubband scattering, etc. The task of the further studies will be
to understand why  the validity of Eq.~(\ref{maineq}) in quantum transport is sample-dependent.

 $K_{ee}^{xx}$ values for the majority of our samples is comparable with those measured by  other groups \cite{magnetron,flakesexample1,DeiEEEvidence,WangEEEvidence,TakagakiEEEvidence}, whereas the temperature dependence of the Hall effect in these papers was not analyzed. However, in Ref.~\cite{HeJAP} [7 nm film shown in Fig.~4(a) and 4(d)], and in Ref.~\cite{Barzola} (Fig.~8),  one can see correlation between the low-$T$ resistivity upturn and increase of the Hall coefficient; remarkably, the resistivity per square for these films and the scale of the Hall coefficient variation  are comparable to those for our samples.

In several papers\cite{WangEEEvidence,DeiEEEvidence}, attempts have been made to  deduce EEI correction from its Zeeman splitting  dependence by applying magnetic field parallel to the sample plane and comparing the data with the classical theory by Lee and Ramakrishnan \cite{LeeRamakrishnan}.
Two notes should be made here: (i) straightforward  application of the Lee and Ramakrishnan theory \cite{LeeRamakrishnan} to chalcogenides is not justified, because even without Zeeman splitting, the electron-electron interaction in multi-component carrier system in topological insulator is not simple \cite{LuShen,konig}. Possible scattering between bulk and surface states and between the opposite surface states will further complicate EEI correction, similarly to the multi-valley system \cite{punnoose, kuntsevich_PRB_2007}. Moreover, in case of a strong spin-orbit coupling, triplet terms of the EEI correction are already suppressed, and additional Zeeman splitting is not expected to affect EEI correction \cite{altshulerreview}. (ii) Experimentally Zeeman effect on the EEI correction in topological insulators can be revealed and disentangled from all other possible orbital effects by temperature dependence of the Hall effect in the presence of Zeeman field. Similar measurements were already performed in two-dimensional system with ordinary spectrum \cite{kuntsevichee}.

\section{Conclusions}

  We have shown that in various thin films of Bi$_2$Te$_3$ and Bi$_2$Se$_3$ topological insulators, the  low temperature  Hall coefficient shows a logarithmic temperature correction. In several samples the low-$T$ behavior of both resistivity and Hall effect was consistently quantitatively explained
 by only two mechanisms: weak anti-localization and electron-electron interaction correction, with the latter being dominant in temperature dependence of the resistivity. We have come to this conclusion  by using only the phenomenological property of the  electron-electron interaction correction  not to affect the Hall  conductivity. Thus, our method is an alternative way to determine the electron electron interaction constant  $K_{\rm ee}$. On the basis of our data it is impossible to say where the EEI correction comes from: topological surface states or bulk.
In our studies, we tested the robustness of the relationship between the corrections to Hall resistance and diagonal resistivity for  samples  with different mobility, substrate, and composition. As a result we found that the agreement with interaction correction theory  is perfect in some cases and imperfect in other. However, we didn't reveal a strong contradiction with theory and this one of  our main findings.

 Our study  poses new questions: Why amplitude of the electron-electron interaction correction to conductivity in high-density low-mobility films is so large? Why in some samples the phenomenological agreement between experimental data and theory of quantum corrections is not perfect? A practical consequence of our results is that the  Hall slope in topological insulator thin films does not exactly correspond to  carrier density: it includes quantum correction that may be as high as few 10$\%$ in low-mobility films and should be taken into account.

\section{Acknowledgements}

We thank I.~S.~Burmistrov and G.M.Minkov for discussions.
The work was supported by RFBR (16-29-03330). VAP also acknowledges the Indian-Russian Grant DST-MSE
14.613.21.0019 (RFMEFI61314X0019) for support. Magnetotransport measurements have been done using research equipment of the LPI  Shared Facility Center.

\section{APPENDIX I. ELECTRON-ELECTRON INTERACTION CORRECTION TO CONDUCTIVITY IN MULTY-COMPONENT SYSTEM.}

In this section we consider conductivity of a multi-component system in the presence of EEI correction.
We assume that the property Eq.~(\ref{property}) for the EEI correction is valid and the classical part of the conductivity tensor is just a sum of the contributions from all components, i.e:
\begin{equation}
\sigma=\sum_{i=1}^N \frac{n_ie\mu_i}{1+(\mu_i B)^2}\left( \begin{array}{cc}
 1  & \mu_i B \\
-\mu_i B  & 1 \\
\end{array} \right) +\left( \begin{array}{cc}
 \Delta\sigma_{ee}  & 0 \\
0  &  \Delta\sigma_{ee} \\
\end{array} \right)
\label{app1}
\end{equation}

Here $N$ is the number of conductive channels. The calculations of the inverse tensor are trivial but very lengthy. We therefore restrict ourselves to two relevant cases. The first case is the low-field limit $B\mu_i\ll 1$ for all $i$. In this limit we can neglect all quadratic in field terms and consider only linear-in-$B$ Hall resistance. Correspondingly, we have:
\begin{equation}
\rho_{xx}=\frac{1}{\sum_{i=1}^N n_ie\mu_i}\cdot\left [ 1 -\frac{\Delta\sigma_{ee}}{\sum_{i=1}^N n_ie\mu_i}\right ]
\end{equation}
\begin{equation}
\rho_{xy}=\frac{\sum_{i=1}^N n_ie\mu_i^2B}{\left (\sum_{i=1}^N n_ie\mu_i\right )^2}\cdot\left [ 1 -\frac{2\Delta\sigma_{ee}}{\sum_{i=1}^N n_ie\mu_i}\right ]
\label{multiliq}
\end{equation}

It is easy to see that Eq.~(\ref{maineq}) is fulfilled  also in the multi-component system. Another case, related to sample Bi$_2$Se$_3$-724, considers high-mobility sample, where condition $\mu B\ll 1$ is not fulfilled. In order to avoid cumbersome formulas we consider $N=2$ and assume a small density $n_1$ of high-mobility ($\mu_1\gg \mu_2$) surface electrons and much larger density $n_2\gg n_1$ of bulk electrons. The linear in field terms of Hall resistance are given by Eq.~(\ref{multiliq}), while for cubic-in-field terms, neglecting $(\mu_2B)^2$ we have:
\begin{equation}
\frac{d^3\rho_{\rm xy}}{6dB^3}=\frac{n_1\mu_1^4}{\sigma_D}\left [ \frac{2\Delta\sigma_{ee}}{\sigma_D}\left ( 1- \frac{n_1\mu_1n_2\mu_2}{\sigma_D^2}\right )-\frac{n_2^2\mu_2^2}{\sigma_D^2}\right ]
\label{cubic}
\end{equation}

This equation means that not only linear in field term in Hall resistivity should slightly vary with temperature, but also cubic-in-field one. Indeed, as it can be obtained from Fig.\ref{fig724}b, the slope of the $\rho_{\rm xy}/B$ versus $B^2$   increases by $\sim 1\%$ as temperature increases from 0.3 K to 30 K. This effect appears to be too small to identify logarithmic dependence, although the sign of the effect agrees with Eq.~(\ref{cubic}). We believe that taking higher order terms and more groups of carries into account one would make the agreement better.

\section{APPENDIX II. ESTIMATES OF ERRORS.}
\begin{figure}
\centerline{\psfig{figure=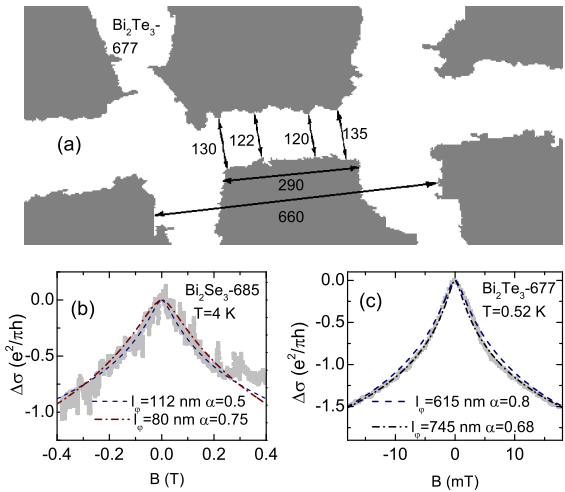, width=220pt}}
\caption{(Color online) a) Digitized photography of the manually scratched sample Bi$_2$Te$_3$-677 with numbers (in micrometers) used to estimate $w/l$ ratio and its error.(b) Weak antilocalization magnetoresistance data and two HLN curves for the noisy sample Bi$_2$Se$_3$-685, the parameters of the curves are indicated on the panel. (c) Weak antilocalization data and two HLN curves for the low-resistive sample Bi$_2$Te$_3$-677, the parameters of the curves are indicated on the panel.}
\label{appendix}
\end{figure}

In our case there are two types of errors: systematic and random.
The systematic ones are related to the uncertainty of the geometry, i.e. to the definition of $w/l$ ratio. In Fig. \ref{appendix}a we demonstrate that the average width fluctuates across the sample, correspondingly conductivity recalculated from the resistance has an uncertainty (about 10-15\% in all scratched samples). As a result both $\alpha$ and $K_{ee}$ has the same geometrical uncertainty ($\Delta \alpha_g/\alpha=\Delta K_{ee\,g}/K_{ee}=\Delta w/w+\Delta l/l$). Nevertheless this type of error does not affect mutual relations between various mechanisms, or between $K_{ee}$, determined from different methods, i.e. it does not affect the main statement of the paper.
Random errors in our case come from several sources:
(1) The errors related to spread of $\alpha$ values at different temperatures, or spread of $\Delta\sigma_{\rm ee}(\ln (T))$ values used for linear interpolation.
(2) For lithographically defined low-mobility samples the resistivity signal was noisy (see e.g., Fig.~\ref{fig3}). Such noise might come from bulk carriers \cite{noise, noise2} or from unstable contacts. Sometimes resistivity exhibited jumps (see e.g. 4\,K and 16\,K curves in Fig.~\ref{fig3}), that makes the overall fitting procedure ill-defined, because the fit is non-linear and depends on symmetrization, definition of $B=0$ resistivity, and the field range used for fitting. The best thing we can do is to play manually with all these parameters, find $\alpha$ and $l_\phi/l$ by minimizing the the standard deviation in each case and see how these parameters are scattered. Fig. \ref{appendix}b illustrates that the noisy data cause large uncertainty in the determined $\alpha$ and $l_\phi/l$ values: indeed to make WAL curve fit visibly distinguishable from the experimental data one has to vary $\alpha$ by 50\% and $l_\phi$ by 30\%. For comparison we show in Fig. \ref{appendix}c the data for much cleaner sample, where variations of $\alpha$ and $l_\phi$ are significantly smaller.

Having  estimated all types of errors we evaluate the error of $\alpha$ or $K_{\rm ee}$ as:
\begin{equation}
\frac{\Delta \alpha}{\alpha}=\sqrt{\sum_i \left(\frac{\Delta\alpha_i}{\alpha}\right )^2} \\
\end{equation}
\begin{equation}
\frac{\Delta K_{\rm ee}}{K_{\rm ee}}=\sqrt{\sum_i \left(\frac{\Delta K_i}{K_{\rm ee}}\right )^2}
\end{equation}

These values are indicated in Table~\ref{summarytable}.

\end{document}